# Analyzing the Impact of Inter Cooperation Region Interference in Coordinated Multi-Point Uplink Networks

S. Alireza Banani and Raviraj S. Adve

*Abstract*—We analyze the uplink of coordinated multi-point (CoMP) networks in which cooperation can be amongst $N = 2$ or $N = 3$ base stations (BSs). We consider a 2-D network of BSs on a regular hexagonal lattice wherein the cooperation tessellates the 2-D plane into cooperation regions (CRs); specifically, we analyze the impact of the interference between the CRs in the network. Our model accounts realistic propagation conditions, particularly including shadowing. We obtain accurate, closed-form, approximations for the user capacity coverage probability (CCP) and the ergodic capacity at each point within the CR. To provide a network-level analysis, we focus on the locations within each CR with the minimum CCP – "the worst-case point(s)". The worst-case CCP and/or ergodic capacity can be used in parametric studies for network design. Here, the analysis is applied to obtain the relationship between cell size and CCP and, thereby, the required density of BSs to achieve a chosen target capacity coverage. The analysis also allows for a comparison between different orders of BS cooperation, quantifying the reduced required BS density from higher orders of cooperation. Comprehensive simulations are used to illustrate the accuracy of our analysis, including the approximations used for analytic tractability.

## I. INTRODUCTION

Coordinated multi-point (CoMP) refers to a range of techniques that enable dynamic coordination between base stations (BSs) in a cellular network in order to suppress inter-cell interference (ICI) and improve overall network performance (e.g., to increase capacity [1],[2]). CoMP techniques can be categorized into joint processing (JP) and coordinated scheduling/beamforming (CS/CB) [3]-[4]. Given a fixed number of antennas per BS, JP-CoMP provides a larger capacity improvement than does the CS/CB approach, since data is shared among multiple points for joint transmission [2]; JP-CoMP is, therefore, the focus of this paper.

Several researchers have investigated the performance of JP-CoMP under many different scenarios. These are too numerous to fully list here, but some representative examples are [5]-[24]. Of these, in [5]-[7] the authors investigate the behavior of JP-CoMP under different precoding methods and perfect channel state information (CSI). The works in [8]-[12] studied the effect of imperfect CSI, limited backhaul and limited feedback on the capacity of CoMP networks. In [13]-[15] the authors analyze a CoMP network in the uplink with a centralized server for joint processing with limited backhaul links; here much of the work is in finding the achievable rate region and the rate gap to an upper bound. The uplink network model with joint processing at the central server is related to, but is different from, JP-CoMP with finite-capacity backhaul links deployed between the cooperating BSs. In this latter setting, the uplink model becomes an interference channel with partial receiver cooperation [16]-[17]. The analysis of such a network model under the Wyner model for BSs locations has been carried out e.g., in [18] considering cooperation strategies such as decode-and-forward and compress-and-forward. In, e.g., [19]-[23], the authors evaluate the performance of CoMP schemes in heterogeneous networks. However, they do not provide a tractable analytic formulation for these networks and the results are based largely on simulations.

Although CoMP networks have been researched for several years, there are modeling aspects which justify further investigation since they can lead to new insights into system design. In a JP-CoMP network, each BS cooperates with its adjacent BSs with each group of cooperating BSs defining a cooperation region (CR). In the previous works e.g., in [5]-[24], the joint processing system implemented across each of the CRs is modeled as a multiple-access channel in the uplink and a broadcast channel in the downlink, giving rise to the concept of a network multiuser multiple-input, multiple-output (MIMO) system [24]; interference between users within the CR can then be suppressed (intra-CR interference).

The focus of this paper is to analyze the impact of inter-CR interference (ICRI), the interference from users in *other co-channel CRs*, on the uplink of a JP-CoMP network. As far as the authors are aware, no one has yet analytically derived the impact of ICRI, the residual interference after the CoMP processing scheme eliminates the intra-CR interference. One important consideration here is that of frequency reuse – if at least one of the cooperating BSs is shared between two co-channel CRs, with significant probability the desired signal gets corrupted by high ICRI. This is because, in an uplink network, the interfering users can be arbitrary close to the

The authors would like to acknowledge the financial support of TELUS and the National Science and Engineering Research Council (NSERC) of Canada.
S. A Banani, and R. S. Adve are with the Dept. of Electrical and Computer Engineering, University of Toronto, ON, Canada; e-mails: alireza.banani@utoronto.ca, rsadve@comm.utoronto.ca

shared BS(s) between two co-channel CRs, resulting in the interference dominating the desired signal. As a result, a suitable resource allocation scheme amongst CRs is required to keep co-channel CRs physically separated.

In this paper, we account for ICRI in a JP-CoMP uplink network wherein BSs are located on a hexagonal grid. Our particular focus is on BS cooperation orders of $N = 2$ or $N = 3$. For these cooperation orders, a resource allocation of frequency reuse-6 amongst different CRs perfectly suits the geometry and can reduce ICRI to some extent. Our model allows for different path loss exponents (PLEs), network parameters, and realistic propagation conditions, particularly shadowing. We first provide, in closed-form, an expression for the average ICRI for the three examples of path loss exponents $\alpha = 3$, $\alpha = 3.5$ and $\alpha = 4$ (non-integer values of the PLE require a single numerical integral). We then obtain accurate analytical approximations for the capacity coverage probability (CCP) and the ergodic capacity at each point of a CR. The CCP is the probability that an individual user can achieve a capacity above a chosen threshold.

Motivated by the desire to provide network design tools, we focus on the location(s) within each CR with the minimum CCP – referred to as "the worst-case point(s)". This work follows the same trend as in, but differs from, our work in [25] wherein ICRI is ignored and ZF receive beamforming is used to suppress interference between users within each CR. The worst-case CCP has been used, e.g. in [26]-[28], for network design in grid-based networks without cooperation amongst BSs. However, in these works, closed-form expressions were achieved only for a restricted case of small number of interfering BSs under shadowing or small-scale fading. For the more general cases, tractable analyses are not available and researchers resort to numerical integration and Monte Carlo simulations.

As a design example, in this paper, the CCP at the worst-case points is used to obtain the required density of BSs (equivalently, the optimal size of the cell in the grid model) to ensure that the achievable capacity at *all points* in the network is above a target capacity with a chosen probability (a target CCP). This corresponds to designing a CoMP network with a coverage probability guarantee. The formulation developed here also allows a comparison between the results of different orders of BS cooperation In particular, we quantify the gains in the reduced required BS density from increasing order of cooperation.

As in other CoMP systems, a significant assumption here is the availability of a high-speed backhaul for information exchange (data, control, synchronization, CSI) between the BSs. We also assume perfect CSI at the BSs. Although, in practice, this information is estimated and quality of these estimations is an important consideration, this issue is a well-trodden path, and is not considered here. In short, while we recognize that there is a significant overhead required to support the capacity that is optimized here, we do not focus on this overhead.

The rest of the paper is organized as follows. Section II describes the JP-CoMP system model. The analysis, that is the central contribution of this paper, is presented in Section III; supporting simulation results are presented in Section IV including simulations to illustrate the accuracy of the approximations required in Section III. Section V summarizes and concludes the paper with some discussion.

The notation used is conventional: matrices are represented using bold upper case and vectors using bold lower case letters; $(\cdot)^H$, and $(\cdot)^T$ denote the conjugate transpose, and transpose, respectively. A vector $\mathbf{a} \sim CN(0,1)$ comprises independent and identically distributed (i.i.d.) zero-mean complex Gaussian random variables, each with unit variance. $Q(x)$ represents the standard $Q$-function, the area under the tail of a standard Gaussian distribution. Finally, $E\{\cdot\}$ denotes expectation.

## II. CoMP Uplink System Model and Inter-CR Interference

### A. Network Geometry

We consider the uplink in an infinite cellular JP-CoMP network where BSs, each with $M$ receive antennas, are arranged according to a hexagonal grid with a hexagon side length of $d$. $N$ adjacent BSs cooperate; this tessellates the service area into CRs with each grouping of $N$ BSs serving $U_N$ users which fall in the corresponding CR. Figs. 1-(a) and 1-(b) show sample CRs for adjacent BSs cooperating with $BS_1$ in a network with cooperation order $N = 2$ and $N = 3$, respectively. In Fig. 1-(a), $CO(1, j)$ denotes the CR associated with $BS_1$ and $BS_j$; similarly in Fig. 1-(b), $CO(1, j, k)$ denotes the cooperation region served by $BS_1$, $BS_j$ and $BS_k$.

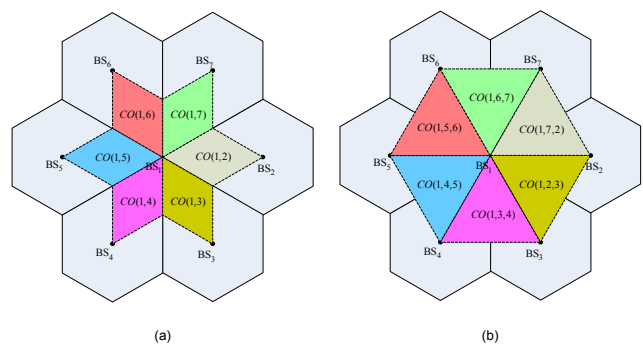

(a)            (b)

Fig. 1 The cooperation regions associated with $BS_1$ for: (a) cooperation order $N = 2$; (b) cooperation order $N = 3$.

With $N = 2$ ($N = 3$) the area of each CR is one-third (half) the area of a hexagonal cell. As a result, in order to make fair comparison between CoMP networks with different orders of BS cooperation, we assume that the number of users within each of the diamond-shaped CRs in Fig.1-(a) is one-third the number of users within a hexagonal cell, i.e., $U_2 = U_1/3$, where $U_1$ denotes the number of users within each hexagon. Similarly, we assume that there are $U_3 = U_1/2$ users within each of the triangular CRs in Fig.1-(b).

We assume that there is no interference amongst users *within* each CR (intra-CR interference) such as due to the use of an orthogonal resource allocation scheme, e.g., different subcarriers in an orthogonal frequency division multiple access (OFDMA) system[1] (the extension of the topic for the case with multiple users coexisting in the same frequency slot and in the same cooperating region, with for example, a multiple-antenna receiver is suggested as future work). The use of OFDMA within each CR has the additional advantage that it enables simple mathematical notation and tractable analysis, as it is sufficient to observe the baseband transmission on a single frequency-flat subcarrier. On each subcarrier, the $N$ BSs cooperatively process data from a single-antenna user (this is not restrictive and the formulation can be modified for users with an arbitrary number of antennas), and form a virtual $NM \times 1$ single-input, multiple-output (SIMO) system. The $NM \times 1$ channel vector $\mathbf{h}_u$ corresponding to $u$-th user can be modeled as

$$[\mathbf{h}_u]_i = \frac{\bar{h}_i^u}{10^{(\mathrm{PL}(r_k^u)+L_k^u)/20}} \quad ; \quad \begin{array}{c} i=(k-1)M+1,\cdots,kM, \\ k=1,\cdots,N \end{array} \quad (1)$$

where $\bar{h}_i^u \sim CN(0,1)$ represents the normalized complex channel gain reflecting small-scale Rayleigh fading and assumed to be statistically independent across users and antennas; $\mathrm{PL}(r_k^u)$ represents the path loss (in dB) from the $u$-th user to the $k$-th cooperating BS, located at a distance $r_k^u$ and can be expressed as $\mathrm{PL}(r_k^u) = 10\alpha \log(r_k^u)$. The effect of shadowing, modeled as a lognormal random variable, is reflected in $L_k^u \sim CN(0,\sigma_L)$, measured in dB; here, $\sigma_L$ denotes the standard deviation of the lognormal fading.

For such a system, on treating the ICRI as additional noise, maximum-ratio-combining (MRC) maximizes the SINR on each subcarrier. Since we focus on the uplink, BSs are the receiving nodes at which the ICRI has to be evaluated. In order to have meaningful communication, co-channel CRs have to be separated such that they do not have any associated BS in common; otherwise, in the case that interfering users are close to the shared BS(s) between co-channel CRs, the ICRI would dominate the desired signal.

Based on the geometry of the CRs shown in Fig.1, we propose the use of frequency-reuse 6 allocation amongst CRs for both the cooperation orders $N=2$ and $N=3$. This is the minimum reuse factor that guarantees that no two co-channel CRs share a BS. Figure 2 illustrates the use of a frequency-reuse 6 allocation amongst CRs in a CoMP network with cooperation order $N=2$ (Fig. 2-(a)) and $N=3$ (Fig. 2-(b)), respectively. In the figure co-channel CRs are labeled with 6 different sets of frequencies $f_1, f_2, \cdots, f_6$.

Interestingly, as is seen from Fig. 2, and emphasized in Fig. 3, each of the cooperating BSs in a CR experiences a common set of nearby co-channel CRs or interfering regions around itself. This is true irrespective of the operating frequency of the CR under consideration. It follows that, for a given PLE $\alpha$ and cooperation order $N$, all the BSs in such a CoMP network experience the same total average ICRI. Figure 3 illustrates the first two tiers of interfering regions as seen by each BS node in the CoMP network. In the figure, the first and second tier of interference regions are labeled as $A_j$ and $B_j$ respectively. We identify the first tier of interference regions as the set of CRs whose distances to the BS node under consideration does not exceeds the distance between two adjacent BSs, i.e., $\sqrt{3}d$ where $d$ is the hexagon side length. Similarly, the second tier of interference regions is identified as the set of CRs whose distances to the BS node under consideration does not exceed $2\sqrt{3}d$.

[1] A CoMP system with cooperation order of $N$ can potentially service $N$ users on the same frequency slot such as by using zero forcing or a minimum mean squared error receiver amongst many multiuser detection schemes. For analytic tractability and to focus on the issue of interference we do not consider this possibility and assume intra-CR interference is eliminated using OFDMA.

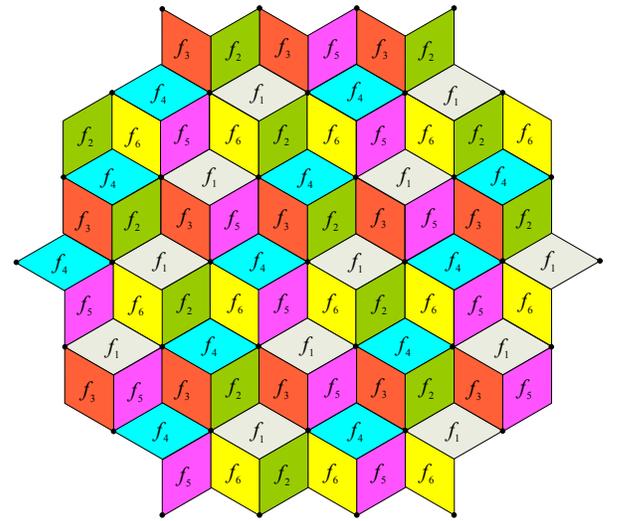

(a)

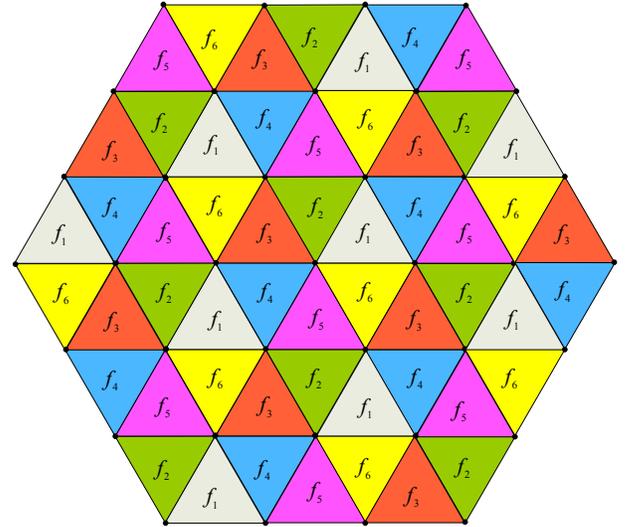

(b)

Fig. 2 The use of frequency-reuse 6 amongst CRs in a CoMP network with: (a) cooperation order $N=2$; (b) cooperation order $N=3$.

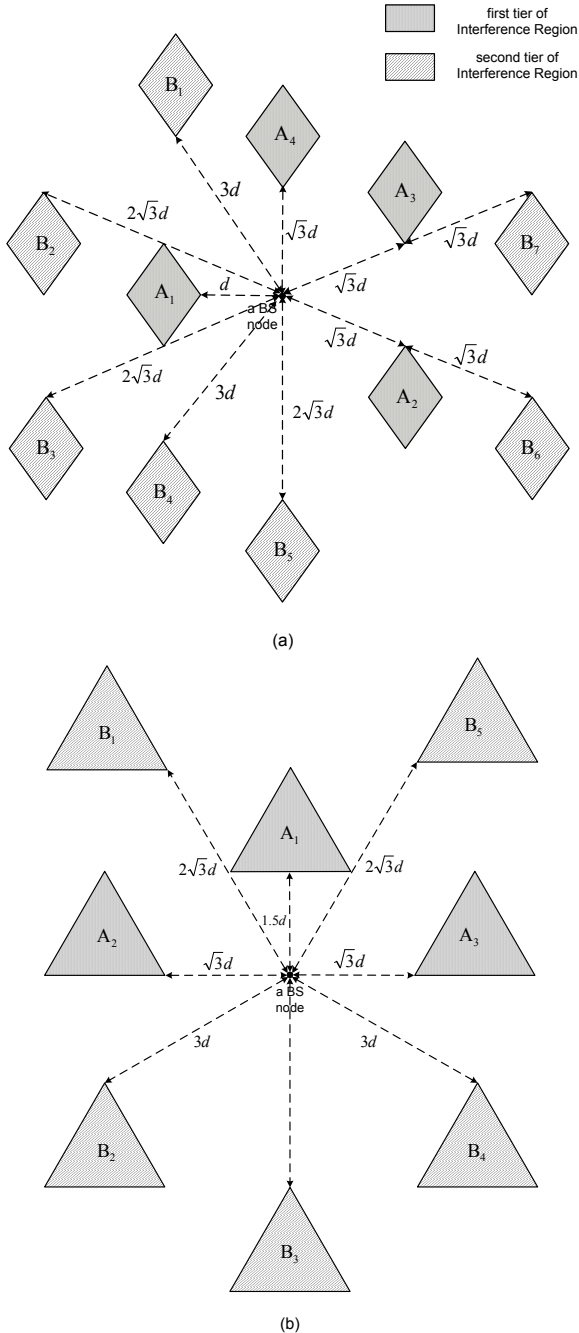

Fig. 3 The first and second tier of interfering regions as seen by each BS node in the CoMP network with frequency reuse-6 amongst CRs for: (a) cooperation order $N = 2$; (b) cooperation order $N = 3$. The first and second tier of interference regions are labeled with $A_j$'s and $B_j$'s, respectively.

Figure 4 compares the result of the interference found through simulating the CoMP network for the PLE of $\alpha = 4$, $M = 1$, user transmit power of $\sigma_s^2 = 20$ dBm and $\sigma_L = 4$ dB. Figure 4-(a) corresponds to a CoMP network with $N = 2$, and Fig. 4-(b) is associated with a CoMP network with $N = 3$. In either network, the interference found from the second tier of interference regions has a negligible contribution to the total interference power (as compared to the interference imposed from first tier only). In particular, with $N = 2$, the interference power from the second tier is 9 dB lower than the interference

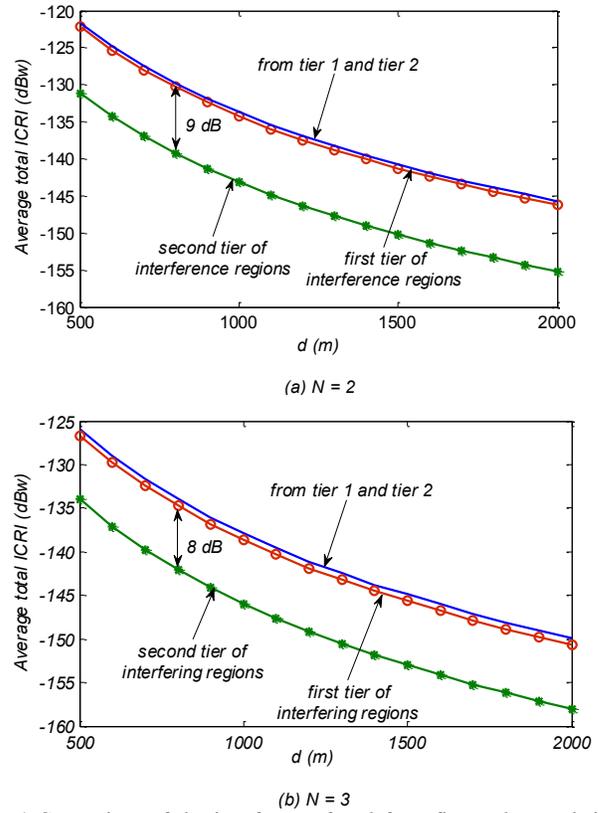

(b) $N = 3$
Fig. 4 Comparison of the interference found from first and second tier of interference regions in a CoMP network with: (a) cooperation order $N = 2$; (b) cooperation order $N = 3$.

power from the first tier. As a result, the first tier of interference regions is the main source of ICRI in the CoMP networks.

Considering one tier of interferers, there are a total of four (three) interfering regions around each BS in a CoMP network when $N = 2$ ($N = 3$). We now analyze the average ICRI, relating it to the hexagon side length $d$ and user transmit power $\sigma_s^2$. The analysis is useful in obtaining analytical expressions for important measures of network performance such as SINR and CCP. With the channel model in (1), using the average of a log-normal random variable, the *average* interference power imposed on a BS at a distance $r$ from a single-antenna user is

$$I^{\text{avg}}(r) = \sigma_s^2 r^{-\alpha} \exp(\sigma_z^2/2), \qquad (2)$$

where $\sigma_z = (0.1\ln 10)\sigma_L$ and the average is over different realizations of small-scale and large-scale fading. In order to obtain the average ICRI imposed from each of the interfering regions, we have to average over the distribution of interfering user distances to the BS which (without loss of generality) is at the origin. To do so, it is more convenient to represent the user location and distances in Cartesian coordinates. Assuming a uniform user distribution inside the CRs, the average ICRI associated with $j$-th interfering region, denoted as $A_j$ in Fig. 3, is given by

$$I_j^{\text{avg}} = \iint_{(x,y) \in A_j} I^{\text{avg}}(x,y) f_{x,y}(x,y) dx dy$$
$$= \sigma_s^2 \exp(\sigma_z^2/2) \iint_{(x,y) \in A_j} (x^2+y^2)^{-\alpha/2} \frac{1}{A_{CR}} dx dy , \quad (3)$$

where $A_{CR}$ is the area of a CR; with $N=2$ we have $A_{CR} = \sqrt{3}d^2/2$, while $A_{CR} = 3\sqrt{3}d^2/4$ for $N=3$. For example for $N=2$, we get

$$I_1^{\text{avg}} = \frac{2\sigma_s^2 \exp(\sigma_z^2/2)}{\sqrt{3}d^2} \left[ \int_d^{3d/2} \int_{-\sqrt{3}(x-d)}^{\sqrt{3}(x-d)} (x^2+y^2)^{-\alpha/2} dy dx + \int_{3d/2}^{2d} \int_{-\sqrt{3}(x-2d)}^{\sqrt{3}(x-2d)} (x^2+y^2)^{-\alpha/2} dy dx \right],$$

$$I_2^{\text{avg}} = I_3^{\text{avg}} = \frac{2\sigma_s^2 \exp(\sigma_z^2/2)}{\sqrt{3}d^2} \times \int_{\sqrt{3}d}^{3\sqrt{3}d/2} \int_{\sqrt{3}(x-\sqrt{3}d)/3}^{\sqrt{3}(x-\sqrt{3}d)/3+d} (x^2+y^2)^{-\alpha/2} dy dx,$$

$$I_4^{\text{avg}} = \frac{4\sigma_s^2 \exp(\sigma_z^2/2)}{\sqrt{3}d^2} \left[ \int_{\sqrt{3}d}^{3\sqrt{3}d/2} \int_0^{\sqrt{3}x/3-d} (x^2+y^2)^{-\alpha/2} dy dx + \int_{3\sqrt{3}d/2}^{2\sqrt{3}d} \int_0^{-\sqrt{3}(x-2\sqrt{3}d)/3} (x^2+y^2)^{-\alpha/2} dy dx \right].$$

The average total ICRI is the summation of individual terms $I_j^{\text{avg}}$, e.g., with $N=2$, $I_{\text{total}}^{\text{avg}} = \sum_{j=1}^4 I_j^{\text{avg}}$; whereas with $N=3$, the summation is over three individual terms.

As seen in Fig. 3, the relative locations of the interfering regions with respect to each BS, are completely identified in the 2-D plane with the distances and the side edges given as multiples of $d$ (the hexagon side length). Thus, the limits of the integrals in (3) are linear functions of $d$. Therefore, for a given PLE $\alpha$ and BS cooperation $N$, the individual terms $I_j^{\text{avg}}$ are obtained in the form of $\sigma_s^2 \exp(\sigma_z^2/2)\beta_j(\alpha,N)d^{-\alpha}$, where $\beta_j(\alpha,N)$ is a coefficient associated with the $j$-th interfering region that depends only on $\alpha$ and $N$. To see this, replace $x$ and $y$ in (3) with $\bar{x}=x/d$ and $\bar{y}=y/d$. Since $A_{CR} \propto d^2$, the integral in (3) results in a term proportional to $d^{-\alpha}$; the constant of proportionality is $\beta_j(\alpha,N)$. Consequently, the average total ICRI, imposed on each of the receive antennas of a BS, can be modeled as

$$\hat{I}_{\text{total}}^{\text{avg}}(d) = \sigma_s^2 \exp(\sigma_z^2/2)\beta(\alpha,N)d^{-\alpha}, \quad (4)$$

where $\beta(\alpha,N) = \sum_j \beta_j(\alpha,N)$. Table 1 provides values of $\beta(\alpha,N)$ corresponding to $N=2$ and $N=3$ and the two examples of PLE $\alpha=3$, $\alpha=3.5$ and $\alpha=4$; for these cases the value of $\beta(\alpha,N)$ is found by analytically evaluating the expression in (3)[2].

Table 1: The coefficients $\beta(\alpha,N)$ in the equation (4) for different values of $N$ and $\alpha$.

|  | $\alpha=3$ | $\alpha=3.5$ | $\alpha=4$ |
|---|---|---|---|
| $N=2$ | $\beta(\alpha,N)=0.57$ | $\beta(\alpha,N)=0.435$ | $\beta(\alpha,N)=0.341$ |
| $N=3$ | $\beta(\alpha,N)=0.257$ | $\beta(\alpha,N)=0.175$ | $\beta(\alpha,N)=0.122$ |

Figure 5 illustrates the accuracy of (4) for the two PLEs $\alpha=3$, $\alpha=3.5$ and $\alpha=4$ with $M=1$, $\sigma_s^2=20$ dBm and $\sigma_L=4$ dB. The figure compares the result of the interference found using (4) and that found through simulating the network under consideration. The simulated results are averaged over many realizations of user locations, small scale fading and large scale fading. As is clear, the analytical results and the results from simulations are indistinguishable for all four cases and for all values of $d$. Furthermore, as expected, due to the associated path loss, for each BS cooperation order, the average total ICRI with $\alpha=4$ is almost 30 dB below that with $\alpha=3$.

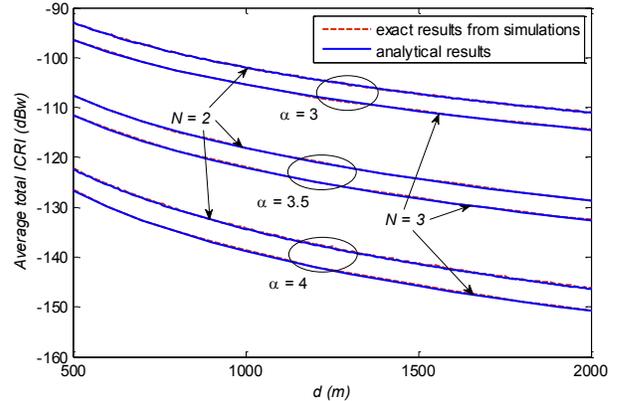

Fig. 5 Average total ICRI (in dBw) imposed on each of the received antennas of a BS as a function of hexagonal cell dimension $d$. The dotted lines are the results from simulations and the solid lines are from the analytical formulation.

### III. CAPACITY COVERAGE PROBABILITY

This section presents the main contributions of this paper. We first derive the CCP and ergodic capacity based on two simplifications: using a lower bound on the CCP and approximating the distribution of a linear combination of lognormal random variables. We defer confirming the accuracy of these approximations to Section IV. We then extend the analysis to evaluate the worst-case CCP and, hence,

---

[2] Evaluating (3) analytically appears to be possible only for cases such as $\alpha=3$ and $\alpha=4$. For other values of the PLE, this integral can be evaluated numerically.

the required density to guarantee a CCP throughout the coverage region.

*A. User CCP and Ergodic capacity within a CR*

Let $n$ denote the frequency-reuse factor amongst CRs. With the total available bandwidth of $W$, equal bandwidth allocation amongst users and $U_1$ users per hexagonal cell, the $u$-th user can achieve rates (in b/s) of $R_2^u = \frac{W}{nU_2}\log_2(1+SINR(u))$ and $R_3^u = \frac{W}{nU_3}\log_2(1+SINR(u))$ with $N=2$ and $N=3$, respectively; here, the superscript '$u$' refers to the $u$-th user in a CR and the subscript reflects the order of cooperation. With our choice of $n=6$ and since $U_2 = U_1/3$ and $U_3 = U_1/2$,, this is equivalent to a user capacity (in b/s/Hz) of $C_2^u = \frac{1}{2}\log_2(1+SINR(u))$ and $C_3^u = \frac{1}{3}\log_2(1+SINR(u))$, respectively. As a result, a general capacity formula for the $u$-th user can be written as $C_N^u = \frac{1}{N}\log_2(1+SINR(u))$ b/s/Hz.

Based on this discussion, the user CCP, defined as the probability that a user can achieve capacity above a chosen threshold, $C_0$, is obtained as

$$\begin{aligned}\mathrm{P}\{C_N^u > C_0\} &= \mathrm{P}\left\{\frac{1}{N}\log_2(1+SINR(u)) > C_0\right\} \\ &= \mathrm{P}\{SINR(u) > \underbrace{2^{NC_0}-1}_{T}\}\end{aligned} \quad (5)$$

The user is therefore in coverage when its achieved SINR is larger than a threshold $T$. In (5), the SINR (using MRC at the cooperating BSs) is given by

$$SINR(u) = \frac{\sigma_s^2 |\mathbf{h}_u^H \mathbf{h}_u|^2}{\mathbf{h}_u^H \mathbf{G} \mathbf{h}_u},$$

where $\mathbf{h}_u$ denotes the channels from user $u$ to the $NM$ receive antennas given in (1). Furthermore, $\mathbf{G}$ is a $NM \times NM$ diagonal matrix with the $i$-th diagonal element $[\mathbf{G}]_{i,i} = \sigma_n^2 + \sigma_s^2 \sum_{j=1}^{N_I} |g_{i,j}|^2$ where $N_I$ represents the total number of co-channel interfering users and $g_{i,j}$ represents the channel (including small-scale fading and log-normal shadowing) from interfering user $j$ to the $i$-th receive antenna of the cooperating BSs under consideration. Also, $\sigma_n^2$ is the variance of the additive Gaussian noise at each receive antenna.

*Approximate SINR Distribution and CCP*: Evaluating the expression in (5) requires averaging over the direct channels, $\mathbf{h}$, and the interfering channels in $\mathbf{G}$. This is extremely complicated and likely intractable. To move forward, in evaluating the CCP we use the average interference in the denominator of the SINR. Thus, the SINR expression is now given as

$$SINR(u) \cong \frac{\sigma_s^2 \mathbf{h}_u^H \mathbf{h}_u}{\sigma_n^2 + \hat{I}_{\text{total}}^{\text{avg}}} = \frac{\sigma_s^2 \mathbf{h}_u^H \mathbf{h}_u}{\sigma_n^2 + \sigma_s^2 \exp(\sigma_z^2/2)\beta(\alpha,N)d^{-\alpha}}, \quad (6)$$

i.e., we replace the instantaneous interference power with the *average* interference power obtained in the previous section. The work in [29] shows that this approach provides a lower bound on the CCP. As we will see in Fig. 5 in the next section, this lower bound is very tight.

The instantaneous achieved SINR[3] in (6) depends on the location of user (via the path loss $\mathrm{PL}(r_k^u)$ in $\mathbf{h}_u$) as well as the instantaneous realizations of the small scale fading $\bar{h}_i^u$ and the lognormal fading $L_k^u$. The achieved $SINR(u)$ can be expressed as

$$\begin{aligned}SINR(u) &= \sum_{k=1}^N 2c\left(\sum_{i=(k-1)M+1}^{kM} \|\bar{h}_i^u\|^2\right)(r_k^u)^{-\alpha}10^{-L_k^u/10} \\ &= \sum_{k=1}^N \underbrace{c\omega_k z_k}_{\xi_k}\end{aligned}, \quad (7)$$

with $c = (1/2)\sigma_s^2/(\sigma_n^2 + \sigma_s^2 \exp(\sigma_z^2/2)\beta(\alpha,N)d^{-\alpha})$, $\omega_k = 2\sum_{i=(k-1)M+1}^{kM} \|\bar{h}_i^u\|^2$, and $z_k = (r_k^u)^{-\alpha}10^{-L_k^u/10}$. Here, $z_k$ is a log-normal random variable with $z_k \sim LN(\mu_{z_k} = -\alpha \ln r_k^u, \sigma_{z_k} = (0.1\ln 10)\sigma_L)$. As the sum of $2M$ independent standard normal random variables, the random variable $\omega_k$ has a chi-squared distribution with $2M$ degrees of freedom ($\omega_k \sim \chi^2(2M)$).

Finally, from $\xi_k = c\omega_k$, it follows that $\xi_k$ has a gamma distribution with parameters $\kappa = 2M/2$ and $\theta = 2c = \sigma_s^2/(\sigma_n^2 + \sigma_s^2 \exp(\sigma_z^2/2)\beta(\alpha,N)d^{-\alpha})$, i.e., $\xi_k \sim \Gamma(\kappa,\theta)$.

From (7), the SINR is a linear combination of the independent lognormal random variables $z_k; k=1,\cdots,N$ with coefficients $\xi_k; k=1,\cdots,N$ that are themselves independent gamma random variables. The key to simplifying this expression is to use the fact that, as shown in [30], for the purposes of outage analyses, linear combinations of lognormal random variables can be closely approximated by a single lognormal random variable. The work in [30] presents several such approximations based on a generalization of the Moment Matching approach. For example, by matching the first and second moments, $SINR(u)$ is approximately distributed as

$$SINR(u) \sim LN(\mu_{SINR(u)}, \sigma_{SINR(u)}), \quad (8)$$

with $\mu_{SINR(u)} = 2\ln(\gamma_1) - 0.5\ln(\gamma_2)$ and $\sigma_{SINR(u)}^2 = -2\ln(\gamma_1) + \ln(\gamma_2)$, where, in turn,

$$\gamma_1 = \sum_{k=1}^N E\{\xi_k\}\exp(\mu_{z_k} + \sigma_{z_k}^2/2), \quad (9)$$

---

[3] Since the SINR in (6) depends on the instantaneous channel realization of the desired user, we will call this the "instantaneous achieved SINR" while acknowledging that this is based on a simplification of the SINR expression.

$$\gamma_2 = \sum_{k=1}^{N} E\{\xi_k^2\} \exp(2\mu_{z_k} + 2\sigma_{z_k}^2)$$
$$+ \sum_{i=1}^{N} \sum_{\substack{j=1 \\ j \neq i}}^{N} E\{\xi_i\} E\{\xi_j\} \exp(\mu_{z_i} + \mu_{z_j} + (\sigma_{z_i}^2 + \sigma_{z_j}^2)/2) \quad (10)$$

The gamma distributed random variables $\xi_k \sim \Gamma(\kappa, \theta)$ have a mean and second order moment of $E\{\xi_k\} = \kappa\theta$ and $E\{\xi_k^2\} = \kappa\theta^2 + \kappa^2\theta^2$, respectively. Therefore,

$$\gamma_1 = \frac{M\sigma_s^2 \exp(\sigma_z^2/2)}{\sigma_n^2 + \sigma_s^2 \exp(\sigma_z^2/2)\beta(\alpha, N)d^{-\alpha}} \sum_{k=1}^{N} (r_k^u)^{-\alpha}, \quad (11)$$

$$\gamma_2 = K\left[(M+1)\exp(\sigma_z^2)\sum_{k=1}^{N}(r_k^u)^{-2\alpha} + M\sum_{i=1}^{N}\sum_{j=1, j\neq i}^{N}(r_i^u)^{-\alpha}(r_j^u)^{-\alpha}\right], (12)$$

with $K = \frac{M\sigma_s^4 \exp(\sigma_z^2)}{(\sigma_n^2 + \sigma_s^2 \exp(\sigma_z^2/2)\beta(\alpha, N)d^{-\alpha})^2}$.

Having found an approximate distribution of the SINR as a lognormal random variable in (8), the CCP in (5) can be easily approximated as:

$$P\{C_N^u > C_0\} = P\{SINR(u) > 2^{NC_0} - 1\}$$
$$\approx Q\left(\frac{\ln(2^{NC_0} - 1) - \mu_{SINR(u)}}{\sigma_{SNR(u)}}\right). \quad (13)$$

The expressions in (11)-(13) lead to an interesting result – if the network is interference limited, due to, for example, a low PLE or extremely small cell sizes, the coverage probability at a given set of normalized distances $\bar{r}_k^u, k = 1,\cdots,N$ is independent of the cell size determined by $d$. This is because, in such networks, the noise variance can be ignored in the denominators of $\gamma_1$ and $\gamma_2$ making them independent of $d$ at a given set of normalized distances $\bar{r}_k^u, k = 1,\cdots,N$. In interference limited networks, therefore, the cell size does not determine coverage probability, a result that is consistent with other system models, e..g, [31], [32]. As an added benefit, in order to analyze an interference-limited CoMP network, it is sufficient to consider a CoMP network with unit cell radius.

*Ergodic Capacity*: Using the fact that for a positive random variable $X$, $E\{X\} = \int_{t>0} P(X>t)dt$, we can use the CCP to approximate the ergodic capacity. An approximation of ergodic capacity is given by

$$C_{ergodic}^u = \int_{C_0>0} P\{C_N^u > C_0\}dC_0$$
$$\cong \int_{C_0>0} Q\left(\frac{\ln(2^{NC_0}) - \mu_{SINR(u)}}{\sigma_{SINR(u)}}\right)dC_0. \quad (14)$$

With the use of change of variable $t \equiv (NC_0 \ln 2 - \mu_{SINR(u)})/\sigma_{SINR(u)}$, we get

$$C_{ergodic}^u \cong \int_{C_0>0} Q\left(\frac{NC_0 \ln 2 - \mu_{SINR(u)}}{\sigma_{SINR(u)}}\right)dC_0$$
$$= \frac{\sigma_{SINR(u)}}{N\ln 2} \int_{-\mu_{SINR(u)}/\sigma_{SINR(u)}}^{\infty} Q(t)dt \quad (15)$$
$$= \frac{\sigma_{SINR(u)}}{N\ln 2}\left[\int_{-\mu_{SINR(u)}/\sigma_{SINR(u)}}^{0} Q(t)dt + \int_{0}^{\infty} Q(t)dt\right]$$

With the use of the identity [33; Eq. (3-58)],

$$\int_{0}^{x} Q(t)dt = xQ(x) + \frac{1}{\sqrt{2\pi}}(1 - e^{-x^2/2}), \quad (16)$$

the ergodic capacity in (15) is obtained as

$$C_{ergodic}^u = \frac{\mu_{SINR(u)}}{N\ln 2} + \frac{\sigma_{SINR(u)}}{N\ln 2}\left[\frac{e^{\frac{-\mu_{SINR(u)}^2}{2\sigma_{SINR(u)}^2}}}{\sqrt{2\pi}} - \frac{\mu_{SINR(u)}}{\sigma_{SINR(u)}}Q\left(\frac{\mu_{SINR(u)}}{\sigma_{SINR(u)}}\right)\right]. \quad (17)$$

The expressions in (13) and (17) are key developments in this paper. We emphasize that both expressions are point-wise in the sense that they provide approximations for the CCP and the ergodic capacity for the $u$-th user at each point in a CR, respectively.

*Sum Ergodic Capacity and Sum CCP*: As an aside, it is worth mentioning that the analysis above allows for expressions for sum capacity as well. Based on (13) and (17), closed-form lower-bound approximations for the sum capacity coverage probability (sum CCP) and sum ergodic capacity can also be found as follows. The achievable sum capacity (in b/s/Hz) within each CR with $U_N$ users is given by,

$$C_N = \sum_{u=1}^{U} C_N^u = \sum_{u=1}^{U} \frac{1}{N}\log_2(1 + SINR(u))$$
$$= \frac{1}{N}\log_2 \prod_{u=1}^{U}(1 + SINR(u)) \quad (18)$$

The sum capacity formula in (18) is lower-bounded by $(1/N)\log_2(1 + \prod_{u=1}^{U}(SINR(u)))$. Therefore, the coverage probability for the sum capacity (a sum CCP) can be approximated as

$$P\{C_N > C_0\} \approx P\left\{\log_2\left(1 + \prod_{u=1}^{U}(SINR(u))\right) > NC_0\right\}$$
$$= P\left\{\prod_{u=1}^{U}(SINR(u)) > 2^{NC_0} - 1\right\} \quad (19)$$

with $SINR(u)$ distributed as in (8). Since we have approximated the SINR as a lognormal random variable, the product in (19) is also a lognormal distribution, i.e., $\prod_{u=1}^{U} SINR(u) \sim LN(\sum_u \mu_{SINR(u)}, (\sum_u \sigma_{SINR(u)}^2)^{1/2})$. As a result, the sum CCP in (19) can be approximated as follows:

$$P\{C_N > C_0\} \approx P\left\{\prod_u SINR(u) > 2^{NC_0} - 1\right\}$$

$$\cong Q\left(\frac{\ln(2^{NC_0} - 1) - \sum_u \mu_{SINR(u)}}{(\sum_u \sigma^2_{SINR(u)})^{1/2}}\right), \quad (20)$$

an expression analogous to (13) for the single-user case. Note that the distribution for the sum ergodic capacity can also be approximated as in (17) with the substitutions, $\mu_{SINR(u)} \to \sum_u \mu_{SINR(u)}$, $\sigma^2_{SINR} \to \sum_u \sigma^2_{SINR}$.

### B. Worst-case CCP within a CR

The analysis in Section III-A is dependent on the location of the user(s) and the CCP expression in (13) can be used to obtain an average CCP throughout the CR as

$$P\{C^u_N > C_0\}_{avg} = \iint_{(x,y) \in A_j} P\{C^u_N > C_0\}_{x,y} f_{x,y}(x,y) dxdy$$

$$= \frac{1}{A_{CR}} \iint_{(x,y) \in A_j} Q\left(\frac{\ln(2^{NC_0} - 1) - \mu_{SINR(u)}(x,y)}{\sigma_{SNR(u)}(x,y)}\right) dxdy$$

where $f_{x,y}(x,y) = 1/A_{CR}$ denotes a uniform user distribution within a CR area; however, this appears to be only possible numerically while obtaining a closed form expression appears intractable. To provide a network level analysis here we use the developments in the previous section to analyze the special case of the worst-case CCP. The choice of worst-case CCP corresponds to a system performance guarantee and has the distinct advantage of analytical tractability. The following proposition sets the location of the worst-case point(s).

*Proposition*: Under the approximations made to derive the CCP above, for the case of $N = 2$ and the associated diamond-shaped CR, the user CCP is minimized at the two edges with $r^{(u)}_1 = r^{(u)}_2 = d$. For $N = 3$, the minimum CCP occurs when the user is located at the centre of the triangular CR ($r^u_1 = r^u_2 = r^u_3 = d$).

*Proof*: See Appendix. ∎

The proposition confirms our intuition that the worst-case point(s) are as far from the cooperating BS as possible. The worst-case CCP is therefore given by (13) with

$$\gamma_1 = \frac{M\sigma^2_s \exp(\sigma^2_z/2)}{\sigma^2_n + \sigma^2_s \exp(\sigma^2_z/2)\beta(\alpha,N)d^{-\alpha}}(Nd^{-\alpha}), \quad (21)$$

$$\gamma_2 = \frac{M\sigma^4_s \exp(\sigma^2_z)N[(M+1)\exp(\sigma^2_z) + M(N-1)]d^{-2\alpha}}{(\sigma^2_n + \sigma^2_s \exp(\sigma^2_z/2)\beta(\alpha,N)d^{-\alpha})^2}. \quad (22)$$

With equal distances to the cooperating BSs, the worst-case results using (21)-(22) are obtained as a function of $d$, or alternatively, the density of BSs in the network via $\lambda = 1/\text{cell Area} = 2/(3\sqrt{3}d^2)$. Thus, for a target value of CCP, the optimal BS density can be obtained such that every point in the CR meets a guarantee on the coverage probability.

In summary, in this section we analyzed the coverage probability in the uplink of a CoMP network with $N$ cooperating BS. The result is based on a lower-bound on the CCP and an approximation for the distribution of the signal power. In the next section we illustrate the accuracy of the approximations and provide some design examples illustrating some applications of the analysis developed here.

It is worth noting that the above analysis including the formulations and the ICRI calculations can be extended for a CoMP network when $N > 3$ BSs cooperate. An overview of how the presented analysis can be extended for a CoMP network with cooperation order $N = 4$ is provided in the Appendix II.

## IV. NUMERICAL RESULTS

The worst-case CCP depends on various parameters of the network such as number of transmit antennas at each BS, the density of BSs, users transmit power and PLE. Thus, our analysis could be used to obtain the various network design parameters to meet a worst-case CCP constraint. In this paper, we obtain the density for BSs when other network parameters are fixed. The parameters in the simulations are: the power of each user is set to $\sigma^2_s \equiv 20$ dBm while the additive noise at each BS antenna is set to $\sigma^2_n \equiv -100$ dBm. Each user is equipped with one receive antenna. In most examples, we define coverage as receiving a capacity of $C_0 = 0.5$ b/s/Hz, i.e., the CCP is given by $P\{C_N > 0.5\}$. The parameters in the simulations are chosen for illustration purposes and choosing other values only scales the results.

In Section III, we needed to use approximations to derive the closed-form expressions for the worst-case CCP. Thus, it is important to first validate the approximations. Figure 6 compares the analytically developed worst-case CCP from the closed-form expressions with the results obtained from Monte Carlo simulations for PLE values of $\alpha = 3$ and $\alpha = 4$. In each system, the number of users per CR is equal to the order of BS cooperation, i.e., $U_N = N$. The cell size is chosen to be $d = 500$ m and the number of antennas per BS is set to $M = 1$. In the simulations we calculate the instantaneous interference caused by each user and not the average as in (4), i.e., the channels $g_{i,j}; i=1,\cdots,NM$, $j=1,\cdots,N_I$ (including path-loss, Rayleigh small-sale fading and log-normal shadowing) are generated setting the standard deviation of the lognormal shadowing to $\sigma_L = 6$ dB.

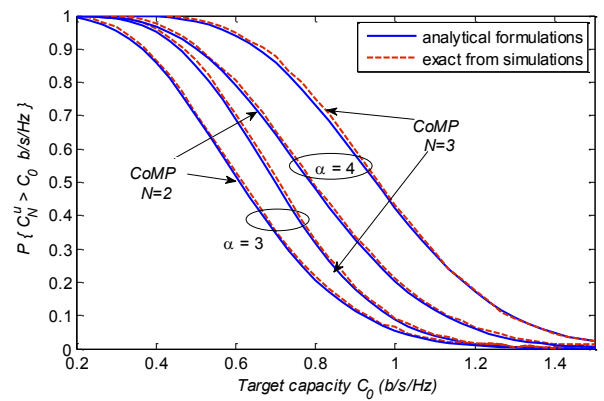

Fig. 6 Comparison of the true and analytic worst-case user CCP as a function of target capacity $C_0$. The dotted lines are the results associated with simulations and the solid lines represent the analytic results in (13).

The simulated curves are obtained by averaging over many realizations of the instantaneous interference. As is clear from the figure, there is a close match between the analytical and simulated results with the difference approaching zero for moderate to high values of target capacity $C_0$. Importantly, the figure illustrates that the approximation to the lower bound on the CCP, used in Section III, is very close and a design based on the expressions derived would be valid.

Figure 7 illustrates the analytic relationship between coverage and cell size. The figure plots the worst–case CCP versus different values of BS density $\lambda$ (equivalently hexagon size $d$ since $\lambda = 2/(3\sqrt{3}d^2)$ ). For either $N=2$ or $N=3$ the CCP is essentially constant for the PLE of $\alpha = 3$. On the other hand, for a PLE of $\alpha = 4$, the CCP is very sensitive to cell size. This is due to the fact that with $\alpha = 3$, as evident from Fig. 5, the ICRI dominates noise and the system is interference-limited; as discussed in Section III, the CCP for an interference-limited system is independent of the cell size. The case of $\alpha = 3.5$ is an intermediate case showing a transition between the two cases.

The BS density at which the network can be treated as interference-limited decreases for larger values of $\alpha$. This is because for a given cell size (BS density) the level of average ICRI decreases with $\alpha$ as seen from Fig. 5. As a result, for smaller range of cell sizes the level of average ICRI remains above the level of noise. Furthermore, for high density networks (small cell sizes) where the networks are interference-limited, the floor of CCP increases with $\alpha$; again, this is due to the ICRI being significantly larger with the lower PLE.

The analysis in Section III allows for multiple antennas at each BS. Figure 8 plots the CCP versus BS density for a fixed PLE of $\alpha = 4$ and $N = 3$, but for different numbers of antennas at the BS, *M*. As a further check on the analysis developed, the dashed lines in the figure represent the results of simulations conducted in a manner similar to that used in Fig. 6. The close match between theory and simulations is again evident. For a chosen CCP value of 0.5 with $M = 1$ the required BS density is $\lambda = 3.8 \times 10^{-7}$ BSs/m² (corresponding to $d \approx 1006$ m). Using $M = 2$ antennas at the BS reduces the required density by almost 40% ($\lambda = 2.3 \times 10^{-7}$ BSs/m²) with further reductions for larger values of *M*. This improvement in CCP is, of course, due to the increase in the SINR at the receiver arising from coherent combining of multiple antennas at the BSs. Furthermore, as expected, the improvement in CCP diminishes as *M* increases.

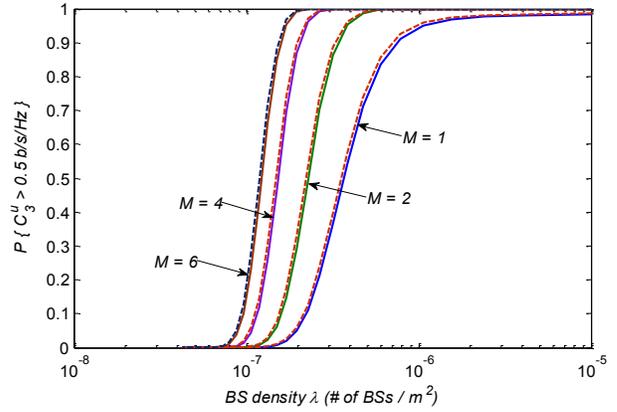

Fig. 8 Worst–case CCP versus BS density for different numbers of antennas at each BS, $N = 3$, in an uplink CoMP network with $N = 3$ and $\alpha = 4$.

In the discussion associated with Fig. 7, we chose a CCP threshold of 0.5, i.e., we require that a user at *any point in the network* is able to achieve a capacity of 0.5 b/s/Hz with probability 0.5. Fig. 9 illustrates that this constraint is, in fact, achieved. The figure plots the CCP in the triangular CR for the case of $N = 3$ and $M = 1$ (corresponding to a density of $\lambda = 3.8 \times 10^{-7}$ BSs/m²). As the figure shows, as required by the analysis, the CCP is always higher than 0.5 and reaches its minimum when user is located at the worst-case point, which, for the triangular CR, is at the centre of the CR. The simulation shows that the minimum achievable CCP is 0.55, which is higher that the target value of 0.5. This is expected, since the analytical results are based on a lower bound on the true CCP results. As a result, the design based on the analytical formulation developed in this paper guarantees maintaining the CCP in the entire CR above the chosen target value.

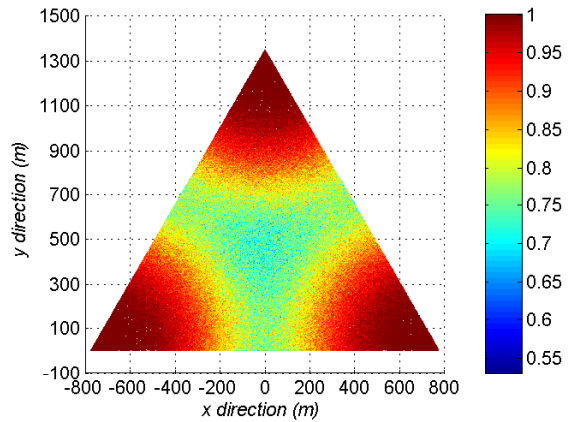

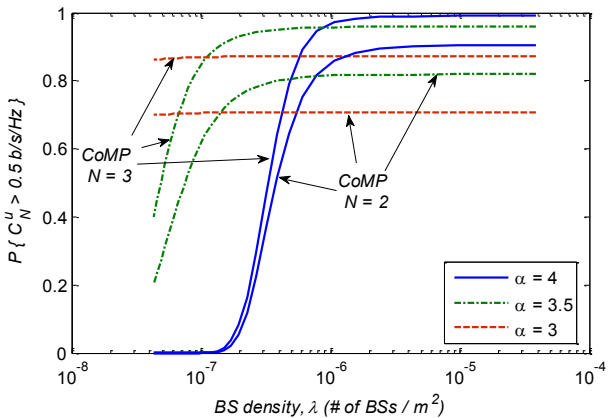

Fig. 7 Worst-case CCP versus BS density for $C_0 = 0.5$ b/s/Hz for different orders of BS cooperation and PLEs.

Fig. 9 User CCP as a function of user location (for the target capacity 0.5 b/s/Hz) within a triangular CR of an uplink CoMP network with $N = 3$, and $d = 1006$ m (corresponding to $\lambda = 3.8 \times 10^{-7}$ BSs/m²).

As noted in Section III.A, the BS density can also be adjusted subject to a target value of worse-case ergodic capacity. Fig. 10 illustrates the corresponding worst-case user ergodic capacity, for different values of BS density at the worst-case point(s) and three values of PLE. For the target capacity of $C_{ergodic}^{u,w} = 0.5$ b/s/Hz with a PLE of $\alpha = 4$ and $N = 2$, the required BS density turns out to be $\lambda = 3.6 \times 10^{-7}$ BSs/m$^2$. This value corresponds to $d = 1034$ m. Figure 11 illustrates the resulting ergodic user capacity within the CR for the case of $N = 2$ (resulting in a diamond-shaped CR). The figure verifies that with the choice of $d = 1034$ m, the minimum capacity of 0.5 bits/s/Hz for each user is achieved in the network. As before, since the design is based on a lower bound, the capacity at the worst-case point is, in fact 0.514 bits/s/Hz.

Finally, we compare networks with different orders of BS cooperation. As a point of comparison, we also simulate the performance of the no-cooperation case (no-CoMP) in which users within each cell are governed by their own BS only. We consider no-CoMP networks with both frequency reuse-1 and reuse-7. In frequency reuse-1, all the hexagonal cells share the same bandwidth and the effect of interference from the users in the surrounding cells has to be accounted for. On the other hand, with frequency reuse-7, the interference from up to two tiers of neighboring cells is eliminated, however the capacity formula has to be multiplied by a pre-log factor of 1/7. For illustration purposes, we consider the simplest case of $M = 1$.

Fig. 12 illustrates the user CCP for the capacity of 0.5 b/s/Hz as a function of BS density for the cases of $N = 1$ (no-CoMP), $N = 2$ and $N = 3$; the plot for the no-CoMP cases is obtained via simulations. As expected, the required BS density decreases with the cooperation order $N$ for a given CCP. However, the greatest gain in the reduced required BS density, is obtained by going from no cooperation ($N = 1$) to CoMP with $N = 2$ and smaller gains are obtained smaller gains are obtained by further increasing $N$ to $N = 3$.

As an added benefit, the formulation developed here allows us to quantify this gain for a given target value of CCP. In particular, for the example of $CCP = 0.5$, the CoMP network with $N = 2$ requires a BS density of 69% as compared to the reuse-1 no-CoMP case; for $N = 3$ this drops to 55%. Compared to a reuse-7 no-CoMP network, the factor in the required BS density is further reduced to 26% and 20% with $N = 2$ and $N = 3$, respectively. Interestingly, for dense networks, the CCP for the reuse-7 no-CoMP network approaches that of the $N = 3$ CoMP network – essentially, for dense networks, the reduction in interference due to frequency-reuse-7 compensates for the pre-log factor of 1/7.

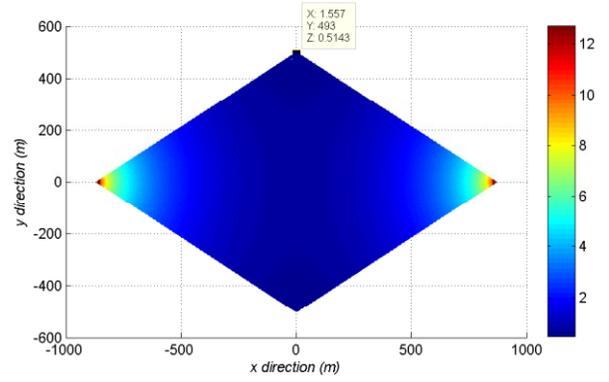

Fig. 11 User ergodic capacity in a diamond-shaped CR of an uplink CoMP network with two BSs cooperating and $M = 1$, $d = 1034$ m; A user is located at different points of the CR.

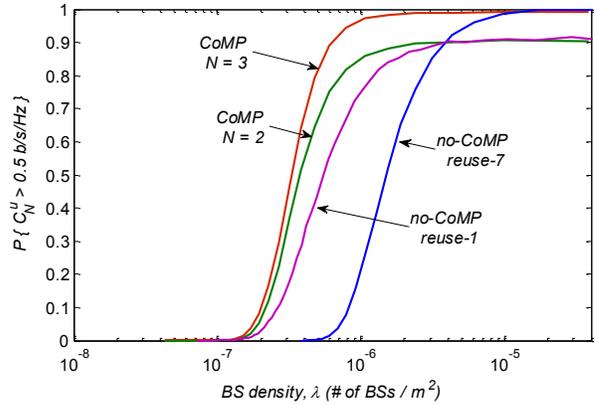

Fig. 12 Comparison of the required BS density in the uplink with different orders of BS cooperation.

## V. SUMMARY AND CONCLUSIONS

This paper analyzes the impact of inter cooperation region interference on the performance of CoMP networks. Specifically, we consider the worst-case performance in the uplink of a CoMP network for two common cases of $N = 2$ and $N = 3$ BSs cooperating; we develop closed-form expressions for the capacity coverage probability. On tessellating the network into cooperation regions, we first developed an accurate closed-form expression for the average inter-cooperation region interference from users in nearby co-channel CRs. From there, we provide accurate closed-form approximations for the CCP (and the ergodic capacity) at each user location in the network. The closed-form expressions are based on analyzing a lower bound on the CCP – and an approximate distribution of the weighted linear sum of log-normal distributions. The simulation results show that this approximation is, in fact, extremely close to the true values.

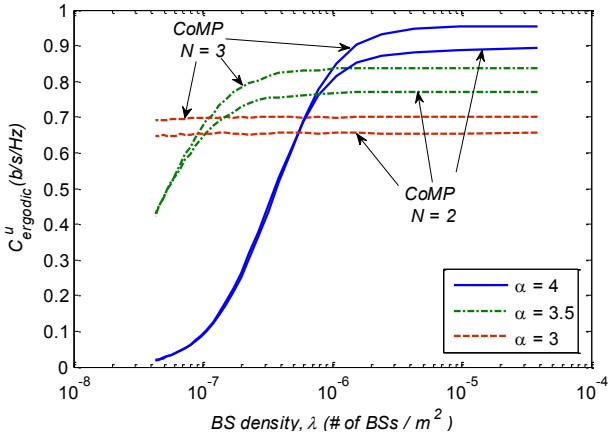

Fig. 10 The worst-case ergodic user capacity versus BS density for different cooperation orders with $M = 1$ under different PLEs.

The presented formulations can incorporate different PLEs and different network parameters, so they can lend themselves to parametric studies for network design. As an example of a parametric design, the worst-case user CCP expression can be used to find the required BS density to maintain the CCP (or ergodic capacity) at all points of the network above a target value. As an added benefit, the formulations allow for comparison between networks with different orders of BS cooperation, quantifying the gain in the reduced required BS density from higher orders of BS cooperation. In particular, for the example of $CCP=0.5$ for the target capacity of 0.5 b/s/Hz, the required BS density in a CoMP network with two BS cooperating is reduced by approximately 30% compared to a reuse-1 network without BS cooperation.

## APPENDIX I

### A. Worst-case point within a CR of a CoMP network with $N=2$

In this subsection we obtain the location(s) of the worst-case point(s) within each CR of an uplink CoMP network with two BS cooperating ($N=2$). Denoting $r_1^w$, $r_2^w$ as the distances corresponding to the worst-case location to $BS_1$ and $BS_2$, respectively, we are to solve the following minimization problem

$$[r_1^w, r_2^w] = \arg\min_{r_1,r_2} P\left\{SINR(u) > \underbrace{2^{NC_0}-1}_{T}\right\}$$
$$= \arg\min_{r_1,r_2} Q\left(\frac{\ln(T)-\mu_{SINR(u)}}{\sigma_{SINR(u)}}\right). \quad (23)$$

In a cellular system, path loss has the most relevant contribution in the degradation of SINR associated with the received signals at each BS. Thus, intuitively, it is clear that the worst-case location is not found near each of the BSs, i.e., $r_1^w$ and $r_2^w$ are relatively large. Correspondingly, in search for $r_1^w$ and $r_2^w$, we may neglect the last term in (12), i.e. $2Me^{-\sigma_z^2}r_1^{-\alpha}r_2^{-\alpha}$, for typical values of $\sigma_z=(0.1\ln10)\sigma_L$ ($2\,\text{dB} \le \sigma_L \le 10\,\text{dB}$ as suggested by [34]) and $\alpha \ge 3$ in practice. As a result, (12) can be well approximated as $\gamma_2 \approx M(M+1)\frac{\sigma_s^4}{\sigma_n^4}e^{(2\sigma_z^2)}(r_1^{-2\alpha}+r_2^{-2\alpha})$. We now get

$$\mu_{SINR} \approx \ln\left[\frac{M^2(\sigma_s^4/\sigma_n^4)10^{-2a/10}e^{\sigma_z^2}(r_1^{-2\alpha}+r_2^{-2\alpha}+2r_1^{-\alpha}r_2^{-\alpha})}{(M(M+1)(\sigma_s^4/\sigma_n^4)10^{-2a/10}e^{2\sigma_z^2}(r_1^{-2\alpha}+r_2^{-2\alpha}))^{1/2}}\right]$$

$$= \ln(B)+\ln\left[(r_1^{-2\alpha}+r_2^{-2\alpha})^{1/2}+\underbrace{\frac{2r_1^{-\alpha}r_2^{-\alpha}}{(r_1^{-2\alpha}+r_2^{-2\alpha})^{1/2}}}_{g(r_1,r_2) \ll (r_1^{-2\alpha}+r_2^{-2\alpha})^{1/2}}\right]$$

$$\approx \ln(B)+\frac{1}{2}\ln(r_1^{-2\alpha}+r_2^{-2\alpha})$$
(24)

with $B=(\sigma_s^4/\sigma_n^4)(M\sqrt{M/(M+1)})$. Also

$$\sigma_{SINR}^2 \approx \ln\left[\frac{M(M+1)(\sigma_s^4/\sigma_n^4)10^{-2a/10}e^{2\sigma_z^2}((r_1^{-\alpha}+r_2^{-\alpha})^2-2r_1^{-\alpha}r_2^{-\alpha})}{M^2(\sigma_s^4/\sigma_n^4)10^{-2a/10}e^{\sigma_z^2}(r_1^{-\alpha}+r_2^{-\alpha})^2}\right]$$

$$= \ln\left[\frac{(M+1)}{M}e^{\sigma_z^2}(1-\underbrace{\frac{2r_1^{-\alpha}r_2^{-\alpha}}{(r_1^{-\alpha}+r_2^{-\alpha})^2}}_{\bar{g}(r_1,r_2) \ll 1})\right] \approx \ln[(M+1)/M]+\sigma_z^2$$
(25)

The last lines in (24) and (25) are obtained by ignoring the associated terms $g(r_1,r_2)$ and $\bar{g}(r_1,r_2)$ that are negligible for large $r_1$, $r_2$ and typical values of $\alpha$. This is justified as follows. In general, at any location, the minimum of $r_1$ and $r_2$ can be related to the larger one, as $\min(r_1,r_2)=\rho\max(r_1,r_2)$ with $0<\rho\le 1$. Without loss of generality, let $r_1 \le r_2$, and so $r_1=\rho r_2$. It follows that $r_2^{-\alpha}=\rho^\alpha r_1^{-\alpha}$. By substituting in $g(r_1,r_2)$ and $\bar{g}(r_1,r_2)$, we get $g(r_1,r_2)=2r_1^{-\alpha}\rho^\alpha/(1+\rho^{2\alpha}) \ll r_1^{-\alpha}(1+\rho^{2\alpha})^{1/2}$, and $\bar{g}(r_1,r_2)=2\rho^\alpha/(1+\rho^\alpha)^2 \ll 1$ which are negligible compared to the associated remaining term in (24) and (25). Now, the minimization problem in (23) can be rewritten as

$$[r_1^w,r_2^w] = \arg\min_{r_1,r_2} Q\left(\frac{\ln(T)-\ln(B)-\frac{1}{2}\ln(r_1^{-2\alpha}+r_2^{-2\alpha})}{\sqrt{\ln[(M+1)/M]+\sigma_z^2}}\right). \quad (26)$$

Since $Q(\cdot)$ is a monotonically decreasing function, the above minimization with respect to $r_1$ and $r_2$ can equivalently be done over the function $\ln(r_1^{-2\alpha}+r_2^{-2\alpha})$ (or minimization over $f(r_1,r_2)=(r_1^{-2\alpha}+r_2^{-2\alpha})$ since $\ln(\cdot)$ is a monotonically increasing function). Consider a diamond-shape CR as shown in Fig. 13.

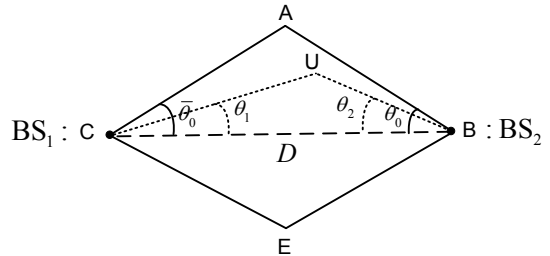

Fig. 13 An example of a diamond-shape CR in a CoMP network with $N=2$

We split the CR in Fig. 13 into two regions (upper triangle ABC and lower triangle BCE). The global minima is the minimum of the results found for the split regions. We note that each point within a triangular region can be identified by the distance pair $[r_1, r_2]$, or the angle pair $[\theta_1, \theta_2]$ where $\theta_i; i=1,2$ is the angle between the side with length $r_i$ and the line connecting the two BSs, or any combinations of distance and angle $[r_i, \theta_j]; i,j=1,2$. The location corresponding to the minimum CCP inside the ABC region, for example, is found as

$$[r_1^w, r_2^w]_{ABC} = \arg \min_{r_1, r_2} P\{C_N^u > C_0\}$$
$$= \arg \min_{r_1, r_2} [f(r_1, r_2) = (r_1^{-2\alpha} + r_2^{-2\alpha})] . \quad (27)$$

For each point inside the triangle ABC, the law of cosines states that $r_1^2 = r_2^2 + D^2 - 2r_2 D \cos\theta_2$, where $D$ is the distance between the two BSs and $\theta_2$ is the angle between BU and BC. Thus the minimization in (27) with respect to $r_1$ and $r_2$ can equivalently be done with respect to $r_2$ and $\theta_2$

$$[r_2^w, \theta_2^w]_{ABC} = \arg \min_{r_2, \theta_2} \left[ \bar{f}(r_2, \theta_2) = \frac{1}{(r_2^2 + D^2 - 2r_2 D\cos\theta_2)^\alpha} + \frac{1}{r_2^{2\alpha}} \right]$$
$$= \arg \min_{r_2} \min_{\theta_2} \left[ \frac{1}{(r_2^2 + D^2 - 2r_2 D\cos\theta_2)^\alpha} + \frac{1}{r_2^{2\alpha}} \mid r_2 \right]$$
$$= \arg \min_{r_2} [\bar{f}(r_2, \theta_2^W = \theta_0) \mid r_2]$$
(28)

where $\theta_0$ is the angle between AB and BC. Equation (28) implies that the worst-case point is located somewhere on AB (i.e, $\theta_2^w = \theta_0$). For each point located on AB, the triangle sine rule states that $r_1/\sin\theta_2^w = r_2/\sin\theta_1$. As a result, the worst-case location on AB is found by searching for $\theta_1$ that minimizes $f(r_1, r_2) = (r_1^{-2\alpha} + r_2^{-2\alpha})$ given $r_1/\sin\theta_2^w = r_2/\sin\theta_1$, i.e.,

$$\theta_1^w = \arg \min_{\theta_1} \left[ \frac{1}{r_1^{2\alpha}} + \frac{1}{(r_1 \sin\theta_1 / \sin\theta_2^W)^{2\alpha}} \right] = \bar{\theta}_0 , \quad (29)$$

where $\bar{\theta}_0$ is the angle between AC and BC. With $\theta_2^w = \theta_0$ and $\theta_1^w = \bar{\theta}_0$, we conclude that the ABC region has the worst-case point located at the point A. Similarly, it can be shown that the worst-case point within the BCE region is located at the point E.

Now that we have the worst-case point(s) associated with different split regions, we identify the one(s) with the worst CCP in the CR. In general, a CR may have more than one location with equal worst CCP. For example, a uniform diamond-shaped cooperation region has two worst-case points located at the two side edges with $r_1^w = r_2^w = d$.

### B. Worst-case point within a CR of a CoMP network with $N = 3$

Similar to the previous case, it can be easily shown that worst-case location to the three cooperating BSs in a CoMP network with $N = 3$ is obtained from the following minimization problem

$$[r_1^w, r_2^w, r_3^w] = \arg \min_{r_1, r_2, r_3} [f(r_1, r_2, r_3) = (r_1^{-2\alpha} + r_2^{-2\alpha} + r_3^{-2\alpha})] , \quad (30)$$

where $r_3^w$ is the distance corresponding to the worst-case location to $BS_3$. With the use of the law of cosines, we can relate $r_1$ and $r_3$ to $r_2$ as $r_1^2 = r_2^2 + D^2 - 2r_2 D\cos\theta_2$ and $r_3^2 = r_2^2 + D^2 - 2r_2 D\cos(\pi/3 - \theta_2)$, respectively (see Fig. 14). Thus the minimization in (30) with respect to $r_1$, $r_2$ and $r_3$ can equivalently be done with respect to $r_2$ and $\theta_2$ by,

$$[r_2^w, \theta_2^w] = \arg \min_{r_2, \theta_2} [\bar{\bar{f}}(r_2, \theta_2)] = \arg \min_{r_2} \min_{\theta_2} [\bar{\bar{f}}(r_2, \theta_2) \mid r_2]$$
$$= \arg \min_{r_2} \left[ \frac{\partial \bar{\bar{f}}(r_2, \theta_2)}{\partial \theta_2} = 0 \mid r_2 \right]$$
(31)

with $\bar{\bar{f}}(r_2, \theta_2) = \frac{1}{(r_2^2 + D^2 - 2r_2 D\cos\theta_2)^\alpha} + \frac{1}{r_2^{2\alpha}}$
$$+ \frac{1}{(r_2^2 + D^2 - 2r_2 D\cos(\pi/3 - \theta_2))^\alpha}$$

From $\partial \bar{\bar{f}}(r_2, \theta_2)/\partial \theta_2 = 0$, the optimal angle is found as $\theta_2^w = \pi/6$ irrespective of $r_2$. As a result, $r_2^w$ is found as

$$r_2^w = \arg \min_{r_2} \left[ \frac{1}{r_2^{2\alpha}} + \frac{2}{(r_2^2 + D^2 - \sqrt{3} r_2 D)^\alpha} \right] = \frac{\sqrt{3}}{3} D = d . \quad (32)$$

The above results indicates that the worst-case point $[r_2^w, \theta_2^w] = [d, \pi/6]$ corresponds to the case when the user is located at the centre of the triangular CR with $r_1^w = r_2^w = r_3^w = d$.

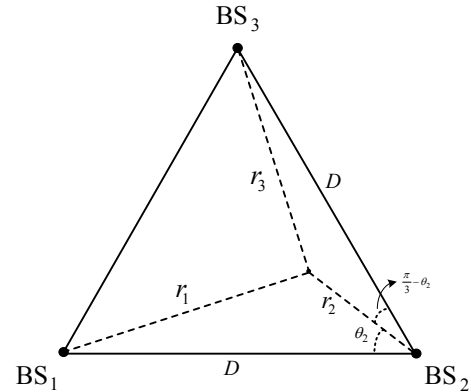

Fig. 14 An example of a triangular-shape CR in a CoMP network with $N = 3$.

### APPENDIX II

This paper has focused on cooperation orders of $N = 2$ and $N = 3$. With the caveat of a more complex interference pattern, cooperation orders of $N > 3$ can also be considered. For example, for the cooperation order $N = 4$, sample CRs for adjacent BSs cooperating with $BS_1$ is shown in Fig. 15. In the figure, $CO(1,j,k,i)$ denotes the CR associated with $BS_1$, $BS_j$, $BS_k$ and $BS_i$. There are the total of 12 disjoint diamond-shaped CRs associated with $BS_1$: 6 CRs with $BS_1$ at a vertex (CRs with solid colors in Fig. 15), plus 6 CRs far apart from $BS_1$ (CRs with dashed colors in Fig. 15). Based on the geometry of the CRs, the minimum frequency reuse factor is equal to the number of CRs having $BS_1$ at one edge, i.e., frequency reuse-6. With the use of frequency reuse-6 it is guaranteed that no two co-channel CRs share a BS at a vertex, keeping the co-channel CRs physically separated from each other; however, intra-CR interference is not eliminated.

Now, the frequency assignment to CRs follows the same as in Fig. 2-(a) for the case of $N=2$. Thus, considering one tier of interferers, there would be a total of four interfering regions around each BS in a CoMP network when $N=4$ (see Fig. 3-(a)). Therefore, the average total ICRI, imposed on each of the receive antennas of a BS, follows the same as in the case of $N=2$. Since the size of the CRs in this case is equal to the size of the diamond-shaped CRs in the case of $N=2$, the user capacity (in b/s/Hz) is obtained as $C_4^u = (1/2)\log_2(1+SINR(u))$, with the SINR term taking into account the cooperation of 4 BSs.

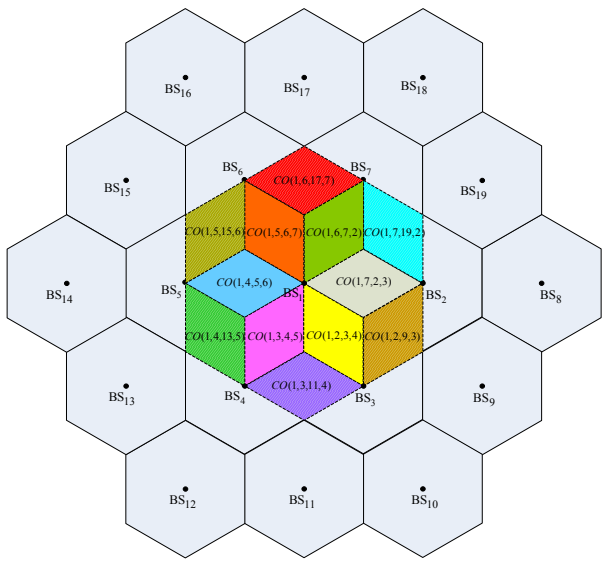

Fig. 15 The cooperation regions associated with $BS_1$ for cooperation order $N=4$

It is worth noting that, higher frequency reuse factor than the minimum one suggested here, can also be used (e.g., frequency reuse-12) to make the co-channel CRs farther apart (resulting in lower ICRI and the elimination of intra-CR interference) at the expense of smaller bandwidth dedicated to each CR.

REFERENCES

[1] R1-083410, "Text proposal for RAN1 TR on LTE-Advanced", NTT DoCoMo.
[2] 3GPP TR36.913, "Requirements for Further Advancements for E-UTRA (LTE-Advanced) (Release 8)," 3GPP, www.3gpp.org, Jun, 2008.
[3] C. Guangquan, S. Mei, Z. Yong, and S. Junde, "Cross-layer adaptation with coordinated scheduling for heterogeneous wireless networks," in Proc. *IEEE Veh. Technol. Conf.* pp. 1-5, Sep. 2010.
[4] F. Minghai, S. Xiaoming, and C. Lan, "Coordinated scheduling based on overload indicator for LTE/LTE-A uplink," in Proc. *IEEE Veh. Technol. Conf.* pp. 1-5, Sep. 2010.
[5] H. Zhang, N. B. Mehta, A. F. Molisch, J. Zhang, and H. Dai, "Asynchronous Interference mitigation in Cooperative Base Station Systems," *IEEE Trans. on Wireless Comm.*, vol. 7, no. I, pp. 155-164, Jan. 2008.
[6] M. H. A. Khan, and M. Ho Lee, "Zero-forcing beamforming with block diagonalization scheme for Coordinated Multi-Point transmission," in Proc. *IEEE Asia-Pacific Conf. Comm.*, pp. 152 – 156, Oct. 2012.
[7] S. A. Banani, Z. Chen, I. Collings, and R. Vaughan, "Point-wise sum capacity maximization in LTE-A coordinated multi-point downlink", in Proc. *IEEE Veh. Technol. Conf.*, pp. 1–5, Jun. 2013.
[8] P. Marsch, and G. Fettweis, "Uplink CoMP under a constrained backhaul and imperfect channel knowledge," *IEEE Trans. on Wireless Comm.* vol. 10, no. 6, pp. 1730-1742, Jun. 2011.
[9] T. Wild, "Comparing downlink coordinated multi-point schemes with imperfect channel knowledge," in Proc. *IEEE Veh. Technol.Conf.*, pp. 1-5, Sep. 2011.
[10] F. Huang, Y. Wang, J. Geng, and D. Yang, "Antenna mismatch and calibration problem in coordinated multi-point transmission system, " *IET Trans. on Comm.*, vol. 6, no. 6, pp. 289-299, Feb. 2012.
[11] S. Han, C. Yang, G. Wang, D. Zhu, and M. Lei, "Coordinated multi-point transmission with non-ideal channel reciprocity," in Proc. *IEEE Wireless Comm. Network. Conf.* pp. 952-957, Apr. 2012.
[12] L. Jingya, A. Papadogiannis, R. Apelfrojd, T. Svensson, and M. Sternad, "Performance evaluation of coordinated multi-point transmission schemes with predicted CSI," in Proc. *IEEE PIMRC Conf.* pp. 1055-1060, Sep. 2012.
[13] L. Zhou, and W. Yu, "Uplink multicell processing with limited backhaul via per-base-station successive interference cancellation," *IEEE J. Sel. Areas Comm.*, vol. 31, no. 10, pp. 1981-1993, Oct. 2013.
[14] A. Sanderovich, O. Somekh, H. V. Poor, and S. Shamai, "Uplink macro diversity of limited backhaul cellular network," *IEEE Trans. on Inf. Theory*, vol. 55, no. 8, pp. 3457–3478, Aug. 2009.
[15] A. del Coso and S. Simoens, "Distributed compression for MIMO coordinated networks with a backhaul constraint," *IEEE Trans. on Wireless Comm.*, vol. 8, no. 9, pp. 4698–4709, Sep. 2009.
[16] I.-H. Wang and D. N. C. Tse, "Interference mitigation through limited receiver cooperation," *IEEE Trans. on Inf. Theory,* vol. 57, no. 5, pp. 2913–2940, May 2011.
[17] C. T. K. Ng, N. Jindal, A. J. Goldsmith, and U. Mitra, "Capacity gain from two-transmitter and two-receiver cooperation," *IEEE Trans. on Inf. Theory*, vol. 53, no. 10, pp. 3822–3827, Oct. 2007.
[18] O. Simeone, O. Somekh, H. Poor, and S. Shamai, "Local base station cooperation via finite-capacity links for the uplink of linear cellular networks," *IEEE Trans. on Inf. Theory*, vol. 55, no. 1, pp. 190–204, Jan. 2009.
[19] M. Shoeb, A. Ibrahim, and H. Elgebaly, "Coordinated multi-point algorithms in pico based heterogeneous networks," in Proc. *IEEE Int. Conf. Advanced Comm. Technol.*, pp. 37-40, Feb. 2012.
[20] L. Falconetti, and. S. Landstrom, "Uplink coordinated multi-point reception in LTE heterogeneous networks," in Proc. *IEEE Int. Symp. Wireless Comm. Systems*, pp. 764-768, Nov. 2011.
[21] N. Miyazaki, X. Wang, and S. Konishi, "A study on homogeneous- and heterogeneous-based additional network

deployments with application of coordinated multi-point operation," in Proc. *IEEE Int. PIMRC Symp.,* pp. 130-135, Sep. 2012.

[22] F. Yuan, C. Yang, G. Wang, and M. Lei, "Adaptive channel feedback for coordinated beamforming in heterogeneous networks," *IEEE Trans. on Wireless Comm.,* vol. 12, no. 8, Aug. 2013.

[23] P. Xia, and C-H. Liu, and J. G. Andrews, "Downlink coordinated multi-point with overhead modeling in heterogeneous cellular networks," *IEEE Trans. on Wireless Comm.* Vol. 12, no. 8, pp. 4025-4037, Aug. 2013.

[24] D. Gesbert, S. Hanly, H. Huang, S. Shamai, O. Simeone, and W. Yu, "Multi-cell MIMO cooperative networks: A new look at interference," *IEEE J. Sel. Areas Comm.,* vol. 28, no. 9, pp. 1380–1408, Dec. 2010.

[25] S. A. Banani, and R. Adve, "Analyzing the reduced required BS density due to CoMP in cellular networks," *IEEE Global Comm. Conf.*, Atlanta, USA, Dec. 2013.

[26] A. J. Goldsmith, *Wireless Communications*. Cambridge University Press, 2005.

[27] T. S. Rappaport, Wireless communications: Principles and practice, Prentice-Hall, 2002.

[28] M. K. Karakayali, G. J. Foschini, and R. A. Valenzuela, "Network coordination for spectrally efficient communications in cellular systems," *IEEE Wireless Comm. Mag.,* vol. 13, no. 4, pp. 56-61, 2006.

[29] B. Hassibi and B. Hochwald, "How much training is needed in multiple-antenna wireless links?", *IEEE Trans. on Inf. Theory*, vol. 49, no. 4, April 2003.

[30] M. Pratesi, F. Santucci, and F. Graziosi, "Generalized moment matching for the linear combination of log-normal RVs: application to outage analysis in wireless systems," *IEEE Trans. on Wireless Comm.*, vol. 5, no. 5, pp. 1122-1132, May 2006.

[31] S. A. Banani, A. Eckford, and R. Adve, "Dependent placement of small cells in a two-layer heterogeneous network with a capacity coverage constraint," submitted to *IEEE Trans. on Comm.* Jan. 2014.

[32] J. G. Andrews, F. Baccelli, and R. K. Ganti, "A tractable approach to coverage and rate in cellular networks," *IEEE Trans. on Comm.*, vol. 59, no. 11, pp. 3122–3134, Nov. 2011.

[33] S. Verdu, *Multiuser detection*, Cambridge university press, 1998.

[34] B. Walke, *Mobile radio networks*, John Wiley & Sons, 2001.